\documentclass[aps,prd,twocolumn,showpacs,10pt,grouppedaddress,superscriptaddress, nofootinbib]{revtex4-1}
\usepackage[utf8]{inputenc}

\usepackage{xcolor}
\definecolor{urlblue}{rgb}{0.2,0.4,0.7}
\definecolor{citegreen}{rgb}{0,0.6,0.2}
\definecolor{linkred}{rgb}{0.9,0.2,0.1}
\usepackage[colorlinks=true, citecolor=citegreen, linkcolor=blue,urlcolor=urlblue]{hyperref}

\usepackage{amssymb}   
\usepackage{amsmath}
\usepackage{diagbox}
\usepackage{dsfont}
\usepackage{accents}
\usepackage{easyReview}
\usepackage{graphics}

\graphicspath{{Figure/}}

\usepackage{amsfonts}
\usepackage{appendix}

\newcommand{\RR}{\texttt{RR}}

\newcommand{\II}{\mathcal{I}_C}
\newcommand{\ep}{\varepsilon}

\allowdisplaybreaks
\begin{document}

\title{NNLO phase-space integrals for semi-inclusive deep-inelastic scattering
}

\author{Taushif Ahmed}
\affiliation{Institute for Theoretical Physics, University of Regensburg, 93040 Regensburg, Germany}

\author{Saurav Goyal}
\affiliation{The Institute of Mathematical Sciences,  Taramani, 600113 Chennai, India}
\affiliation{Homi Bhabha National Institute, Training School Complex, Anushakti Nagar, Mumbai 400094, India}

\author{Syed Mehedi Hasan}
\affiliation{Institute for Theoretical Physics, University of Regensburg, 93040 Regensburg, Germany}

\author{Roman N. Lee}
\affiliation{Budker Institute of Nuclear Physics, Novosibirsk 630090, Russia}
\author{Sven-Olaf Moch}
\affiliation{II.~Institute for Theoretical Physics, Hamburg University, D-22761 Hamburg, Germany}

\author{Vaibhav Pathak}
\affiliation{The Institute of Mathematical Sciences,  Taramani, 600113 Chennai, India}
\affiliation{Homi Bhabha National Institute, Training School Complex, Anushakti Nagar, Mumbai 400094, India}

\author{Narayan Rana}
\affiliation{School of Physical Sciences, National Institute of Science Education and Research, 752050 Jatni, India}
\affiliation{Homi Bhabha National Institute, Training School Complex, Anushakti Nagar, Mumbai 400094, India}
\author{Andreas Rapakoulias}
\affiliation{Institute for Theoretical Physics, University of Regensburg, 93040 Regensburg, Germany}

\author{V. Ravindran}
\affiliation{The Institute of Mathematical Sciences,  Taramani, 600113 Chennai, India}
\affiliation{Homi Bhabha National Institute, Training School Complex, Anushakti Nagar, Mumbai 400094, India}

\begin{abstract}
	We evaluate the phase-space integrals that arise in double real emission diagrams for semi-inclusive deep-inelastic scattering at next-to-next-to-leading order (NNLO) in QCD. Utilizing the reverse unitarity technique, we convert these integrals into loop integrals, allowing us to employ integration-by-parts identities and reduce them to a set of master integrals. The master integrals are then solved using the method of differential equations and expressed in terms of Goncharov polylogarithms. By examining the series expansion in the dimensional regulator, we discover additional relations among some of the master integrals.
	As an alternative approach, we solve the master integrals by decomposing them into angular and radial components. The angular parts are evaluated using Mellin-Barnes representation, while special attention is given to the singular structures of the radial integrals to handle them accurately. Here the results are provided in terms of one-fold integrals over classical polylogarithms. This approach provides a clearer understanding of the origin of soft and collinear singularities.
\end{abstract}

\maketitle

\section{Introduction}
\label{sec:intro}
Semi-inclusive deep-inelastic scattering (SIDIS) stands as a cornerstone in nuclear and particle physics, offering unparalleled insights into the structure and dynamics of nucleons. Unlike inclusive deep-inelastic scattering (DIS), which solely measures the scattered lepton, SIDIS involves the detection of both the scattered lepton and at least one of the produced hadrons. This dual detection capability enriches the data obtained, enabling far more detailed and comprehensive exploration of the internal parton structure and their three-dimensional momentum distribution of nucleons. In addition, this also provides critical insights into how partons combine and fragment to form the observed hadrons. This makes SIDIS a very important process in the upcoming Electron-Ion Collider (EIC) at Brookhaven~\cite{Aschenauer:2019kzf}. The transition from parton to hadron is a non-perturbative process that can be described by fragmentation functions (FFs). These functions quantify the likelihood of a parton fragmenting into a hadron, with the hadron inheriting a certain fraction of the parton's momentum. Production cross sections for various hadron species have been measured in electron-positron, lepton-hadron, and hadron-hadron collisions. Incorporating these data sets into a global fit of FFs requires knowledge of the corresponding parton-level coefficient functions (CFs), which are differential in the momentum of the fragmenting parton. The CFs for hadron-hadron~\cite{Aversa:1988vb} and electron-positron~\cite{Rijken:1996ns,Mitov:2006ic} collisions are currently known up to next-to-leading order (NLO) and next-to-next-to-leading order (NNLO) in QCD, respectively. Very recently, some of us have undertaken the monumental task of computing the CFs analytically for lepton-hadron collisions to NNLO~\cite{Goyal:2023zdi,Goyal:2024tmo}, in parallel with another research group~\cite{Bonino:2024qbh,Bonino:2024wgg}. The NLO CFs for the same are available in refs.~\cite{Altarelli:1979kv,Baier:1979sp,deFlorian:1997zj}. In refs.~\cite{Abele:2021nyo,Abele:2022wuy}, the threshold-enhanced resummed results for some partonic channels have been computed to NNLO and next-to-NNLO (N$^3$LO).

The NNLO QCD corrections for SIDIS involve the evaluation of Feynman diagrams featuring pure two-loop virtual, real-virtual and double-real emissions. While the two-loop virtual diagrams have been present in the literature for quite some time, the remaining ones have been evaluated in refs.~\cite{Goyal:2023zdi,Goyal:2024tmo}.
The real-virtual processes involve one-loop and two-body phase-space integrals, while double real emission processes require three-body phase-space integrals. These integrals are computed with a constraint arising from the requirement of detecting the final state hadron for SIDIS. From the point of view of integration-by-parts (IBP) reduction, this constraint leads to an additional cut propagator, which adds technical complexity compared to fully inclusive processes.

The computation of master integrals (MIs) was one of the primary difficulties in achieving the NNLO calculation. In this article,\footnote{email addresses: \texttt{taushif.ahmed@ur.de}, \texttt{sauravg@imsc.res.in}, \texttt{Syed-Mehedi.Hasan@ur.de}, \texttt{r.n.lee@inp.nsk.su}, \texttt{sven-olaf.moch@desy.de}, \texttt{vaibhavp@imsc.res.in}, \texttt{narayan.rana@niser.ac.in}, \texttt{andreas.rapakoulias@ur.de}, \texttt{ravindra@imsc.res.in}.} we present the detailed computation of the phase-space integrals that arise from the double real emission employing two different methods, namely, the method of differential equations and the radial-angular decomposition.

In the framework of the differential equations method \cite{Kotikov:1990kg,Remiddi:1997ny}, we reduce the differential systems to $\varepsilon$-form \cite{Henn2013, Lee2015} using algorithm of ref.~\cite{Lee2015} as implemented in \texttt{Libra} package~\cite{Lee2021}. We put boundary conditions in the double limit $x\to 1$, $z\to 1$. In this limit it appears to be sufficient to calculate only two constants, the first one expressed via $\Gamma$-functions, the other one via $_3F_2$ hypergeometric function. The $\varepsilon$-expansion of these constants can be made up to arbitrarily high power of $\varepsilon$. The same is also true for the canonical integrals. Therefore, the differential equations approach provides a simple and systematic way to obtain the $\varepsilon$-expansion  of the MIs. Besides, passing to canonical basis, we reveal two nontrivial relations between the MIs, which do not seem to be a direct consequence of the IBP relations.

In an alternative approach, we decompose the phase-space integrals into their radial and angular components. The angular component, being independent of the specific process, appears in various scattering scenarios. Several methods are available for evaluating this component~\cite{vanNeerven:1985xr,Somogyi:2011ir,Lyubovitskij:2021ges}; in this work, we present an approach based on its Mellin-Barnes (MB) representation~\cite{Dubovyk:2022obc}. The angular integrals in SIDIS contain at most two denominators, while integrals with more denominators can be reduced to a linear combination of integrals with fewer denominators using partial fractioning. These integrals are known to be expressible in terms of hypergeometric, Appell, and Fox H-functions, depending on whether the relevant momenta are massive or massless. While the Laurent expansion of hypergeometric and Appell functions in the dimensional regulator can yield the desired results, expanding H-functions in this way is often non-trivial. Moreover, for integrals involving more than two denominators, expressing them as H-functions of several variables makes expansion in powers of the dimensional regulator quite challenging. These difficulties motivate us to adopt an alternative approach based on the MB representation~\cite{Ahmed:2024pxr}.
Using this approach, we derive multifold MB integrals for the angular component, which are then solved analytically in terms of Goncharov polylogarithms. A detailed presentation of this method is available in another publication by some of us in ref.~\cite{Ahmed:2024pxr} where the angular integral involving three denominators is computed for the first time. The latter is also addressed using the differential equation method in ref.~\cite{Haug:2024yfi}.
Combining the angular component with the parametric component requires careful handling of soft singularities, which we illustrate with an example. This approach provides a deeper understanding of the underlying soft and collinear singularities. The complete results are provided in the form of one-fold integrals of classical polylogarithms.

The article is organized as follows: in section~\ref{sec:kin}, we discuss the basic kinematics of the SIDIS process and introduce the concept of constrained phase space. We present the MIs that arise from the double real emission in the computation of the NNLO SIDIS cross section in QCD.
Section~\ref{sec:de} covers the differential equation method, where we reduce the differential systems for MIs to $\varepsilon$-form and express the final results via generalized polylogarithms (GPLs).
In section~\ref{sec:rad-ang}, we demonstrate the angular-radial decomposition method, solving the angular part using MB techniques. We address the treatment of divergences to properly combine the parametric and angular components. Our results in terms of one-fold integrals over classical polylogarithms are provided in ancillary files. This is sufficient for all phenomenological purposes.


\section{Semi-inclusive deep-inelastic scattering}
\label{sec:kin}

We consider the production of a hadron $H$ in SIDIS, where a lepton $ l(p_l) $ scatters off a nucleon $p(P)$, resulting in a final state consisting of the scattered lepton $l(p_l^\prime) $, the observed hadron $ H(P_H) $, and other inclusive final state radiation $ X $:
\begin{align}
l(p_l) + p(P) \rightarrow l(p_l^\prime) + H(P_H) + X.
\end{align}
Here, $ p_l $, $P$, $p_l^\prime$, and $p_H$ denote the respective momenta of the lepton, nucleon, scattered lepton, and hadron.
In SIDIS, the hadron is observed in the final state, while other radiation (denoted by $ X $) is unobserved. The space-like momentum  transfer between leptons, $q=p_l-p_l^\prime$, denotes the momentum of the exchanged virtual vector boson and we set $Q^2=-q^2$. Let  $p_1$ and $p_2$ denote the momenta of initial and final state partons, respectively. We define two dimensionless variables by
\begin{gather}
	x^\prime=\frac{Q^2}{2 p_1\cdot q},\quad z^\prime=\frac{p_1 \cdot p_2}{p_1 \cdot q},
\end{gather}
and set  $Q^2=1$ for convenience. The dependence of the final results on $Q^2$ can be restored on dimensional ground. The physical region is defined by the inequalities
\begin{equation}
	0<x',z'<1\,.
\end{equation}

The triple-differential cross-section for SIDIS can be parametrized in terms of transverse and longitudinal form factors. These form factors are expressed as a sum over all partonic channels, involving the convolution of the parton distribution function (PDF), FF, and partonic cross section.
This computation includes two-loop virtual corrections for the processes $\gamma^* + q(\bar{q}) \rightarrow q(\bar{q})$, one-loop contributions for single-gluon emission $\gamma^* + q(\bar{q}) \rightarrow q(\bar{q}) + g$, and contributions from double real emissions. The quantity $\gamma^*$, $q (\bar q)$ and $g$ denote virtual photon, quark (anti-quark) and gluon, respectively.

Double real emission processes involve three-body phase-space integrals constrained by a delta function, $ \delta\left(z' - \tfrac{p_{1} \cdot p_{2}}{p_{1} \cdot q}\right)$. This additional constraint significantly complicates the computation of these integrals. Fortunately, the argument of the $\delta$-function depends linearly on scalar products involving integration momenta $k_1,k_2,p_2$ and thus can be introduced in the IBP reduction as a cut propagator. We use \texttt{LiteRed2} \cite{Lee2013a,LeeLiteRed2} to reduce these integrals to a set of 20 MIs. The primary objective of this article is to present the detailed computation of these MIs.

\begin{figure}[h]
	\centering
	\includegraphics[width=0.9\textwidth]{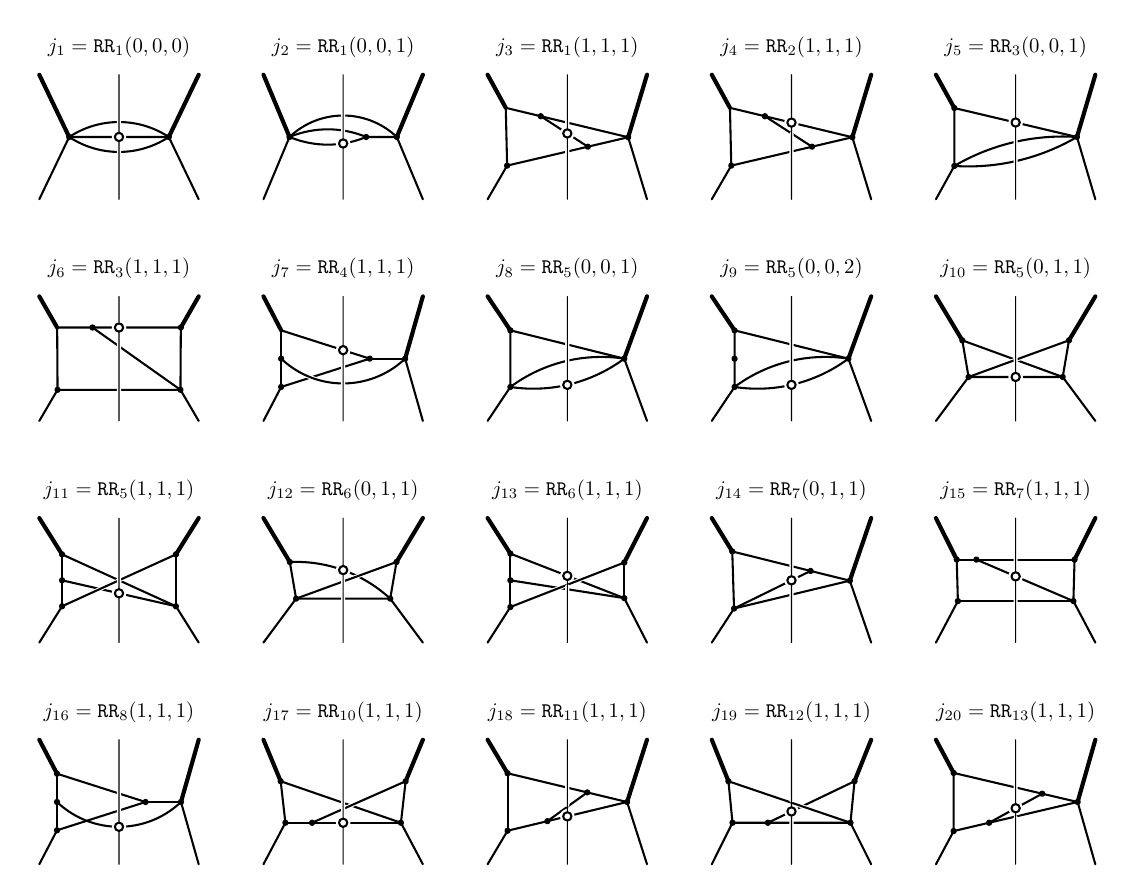}
	\caption{MIs of families $\RR_{1},\ldots, \RR_{13}$. (Cut) lines denote (cut) massless propagators, empty circle marks propagator carrying $p_b$ momentum.}
	\label{fig:mis}
\end{figure}

We define 13 integral families $\RR_{1},\ldots, \RR_{13}$ of two-loop integrals
\begin{align}
	\RR_k\left(n_1,n_2,n_3\right) = e^{-\varepsilon\gamma_{E}}(4\pi)^d\int \frac{[dPS_3]_{z'}}{ D_{N_{k1}}^{n_1}D_{N_{k2}}^{n_2}D_{N_{k3}}^{n_3}}\,,
\end{align}
where $\gamma_{E}=0.577\ldots$ is Euler constant and the indices $N_{ki}$ are defined from the table
\begin{center}
	\begin{tabular}{|c||c|c|c|c|c|c|c|c|c|c|c|c|c|}
	\hline
	\diagbox[height=1.25\line,innerleftsep=1pt,innerrightsep=1pt]{$i$}{$k$}&
	1&2&3&4&5&6&7&8&9&10&11&12&13\\
	\hline
	1&
	1&1&1&
	1&1&
	1&1&1&1&2&2&2&2\\
	2&
	5&5&5&
	6&2&
	2&4&4&4&4&5&5&5\\
	3&
	6&7&8&
	8&4&
	8&5&7&8&5&7&8&6\\
	\hline
\end{tabular}
\vspace{5mm}
\end{center}
and functions $D_1,\ldots, D_8$ read
	\begin{gather}
	D_1=(k_1-p_1)^2\,,\quad
	D_2 = (k_1-q)^2\,,\quad
	D_3 = (k_2-p_1)^2\,,\nonumber\\
	D_4 = (k_2-q)^2\,,\quad
	D_5 = (k_1-p_1-q)^2\,,\nonumber\\
	D_6 = (k_2-p_1-q)^2\,,\quad
	D_7 = (k_1+k_2)^2\,,\nonumber\\
	D_8 = (k_1+k_2-p_1)^2\,.
	\label{eq:D_i}
\end{gather}
The constrained 3-particle phase space reads
\begin{multline}
\label{eq:ph-sp-el}
	[dPS_3]_{z'} = dPS_3 \delta\left(z'- \tfrac{p_{1}\cdot p_{2}}{p_{1}\cdot q}\right)\\
	 =\frac{d^d k_1 d^d k_2 d^{d}p_2}{(2 \pi)^{3d-3}}\delta_+(k_1^2)\delta_+(k_2^2)\delta_+(p_2^2)\\
	\times(2\pi)^d
	\delta^{d}(p_2+k_1+k_2-p_1-q) \delta\left(z'- \tfrac{p_{1}\cdot p_{2}}{p_{1}\cdot q}\right).
\end{multline}
We denote the space-time dimension by $d=4+\varepsilon$, the momenta of emitted partons by $k_1$ and $k_2$. Multiplication by the additional delta function arises from the necessity of detecting an additional particle in SIDIS.
The MIs are depicted in Fig.~\ref{fig:mis}.
\onecolumngrid

\twocolumngrid
Note that in order to find the extra rule for the $\RR_5(-1,0,1)$ integral, we had to use the IBP identities from the supersectors. In the following sections, we solve these integrals using two distinct approaches: the method of differential equations and the method of angular-radial decomposition.

\section{Calculating master integrals using differential equation method}
\label{sec:de}
We construct the differential systems for the MIs with respect to $x'$ and $z'$ and reduce them to $\ep$-form using \texttt{Libra}~\cite{Lee2021}. In order to do this, we introduce the following notations:
\begin{align}
\label{eq:x1x2z1z2}
	x_1 = \sqrt{x'},\quad
	x_2 = \frac{1}{2z'}\left(\sqrt{(1-z')^2+4x'z'}-1+z'\right),
	\nonumber\\
	z_1 = \sqrt{z'},\quad
	z_2 =  \frac{1}{2x'}\left(1+x'-\sqrt{(1+x')^2-4x'z'}\right)\,.
\end{align}
The canonical basis $\boldsymbol{J}$
satisfies
\begin{equation}
	d\boldsymbol{J}=\ep\sum_{i=1}^{16} S_i \omega_i \boldsymbol{J},
\end{equation}
where
\begin{widetext}
\begin{align}
	\omega_1&=d\ln {x'},\quad
	&\omega_5&=d\ln (1-x'),\quad
	&\omega_9&=d\ln (x'+z'),
	&\omega_{13}&=d\ln {\frac{\bar{x}+x' z_2}{z_2}},\quad
	\nonumber\\
	\omega_2&=d\ln {z'},\quad
	&\omega_6&=d\ln (1+x'),\quad
	&\omega_{10}&=d\ln (1+x' z'),\quad
	&\omega_{14}&=d\ln {\frac{\bar{z}+x_2 z'}{x_2}},
	\nonumber\\
	\omega_3&=d\arctan{\frac{z_1}{x_1}},\quad
	&\omega_7&=d\ln (1-z'),\quad
	&\omega_{11}&=d\ln (\bar{x}+2 x' z_2),\quad
	&\omega_{15}&=d\ln {\frac{1+x' z_2}{1-z_2}},\quad
	\nonumber\\
	\omega_4&=d\arctan(z_1 x_1),
	&\omega_8&=d\ln (x'-z'),\quad
	&\omega_{12}&=d\ln (\bar{z}+2 x_2 z'),
	&\omega_{16}&=d\ln {\frac{1+x_2 z'}{1-x_2}}\,,
\end{align}
\end{widetext}
with
$\bar{x}=1-x'$, $\bar{z}=1-z'$, and $S_i$ being some numerical matrices.

\subsection{Boundary conditions}
We fix the boundary conditions by considering the asymptotics $x'\to 1$, $z'\to 1$. The limit $x'\to 1$ corresponds to $s=(p_1+q)^2\to 0$, while the limit $z'\to 1$ corresponds to the maximal energy fraction of the detected final parton.

Using \texttt{Libra}, we find that, in order to fix the boundary conditions, we need to calculate the following asymptotic coefficients:
\begin{gather}
\left[j_1\right]_{\bar{x}^{1+\ep } \bar{z}^{1+\ep }},\ \left[j_2\right]_{\bar{x}^{\ep } \bar{z}^0},\ \left[j_3\right]_{\bar{x}^{-1+\ep } \bar{z}^0},\ \left[j_4\right]_{\bar{x}^{-1+\ep } \bar{z}^{-1+\ep }},
\nonumber\\
\left[j_5\right]_{\bar{x}^0 \bar{z}^{\ep }},\ \left[j_6\right]_{\bar{x}^0 \bar{z}^{-1+\ep }},\ \left[j_7\right]_{\bar{x}^0 \bar{z}^0},\ \left[j_7\right]_{\bar{x}^0 \bar{z}^{1+\ep }},
\nonumber\\
\left[j_9\right]_{\bar{x}^0 \bar{z}^0},\ \left[j_{10}\right]_{\bar{x}^0 \bar{z}^0},\ \left[j_{11}\right]_{\bar{x}^0 \bar{z}^{-1+\ep }},\ \left[j_{12}\right]_{\bar{x}^0 \bar{z}^{-\frac{1}{2}+\frac{\ep }{2}}},
\nonumber\\
\left[j_{13}\right]_{\bar{x}^0 \bar{z}^{-1+\ep }},\ \left[j_{14}\right]_{\bar{x}^0 \bar{z}^0},\ \left[j_{15}\right]_{\bar{x}^0 \bar{z}^0},\
\left[j_{16}\right]_{\bar{x}^0 \bar{z}^{-1+\ep }},
\nonumber\\
\left[j_{17}\right]_{\bar{x}^{-1+\ep } \bar{z}^0},\ \left[j_{18}\right]_{\bar{x}^{-1+\ep } \bar{z}^0},\ \left[j_{19}\right]_{\bar{x}^{-1+\ep } \bar{z}^0},\ \left[j_{20}\right]_{\bar{x}^{-1+\ep } \bar{z}^0},
\end{gather}
where $\left[j_n\right]_{\bar{x}^a \bar{z}^b}$ denotes the coefficient in the asymptotics of $j_n$ in front of $\bar{x}^a \bar{z}^b$ (recall $\bar{x}=1-x'$, $\bar{z}=1-z'$).
It is easy to establish that there is only one region in the limit $\bar{x},\bar{z}\to 0$ (and that the limits are commuting) which contribution reads $\left(\bar{x}\bar{z}\right)^{\ep}\bar{x}^n\bar{z}^m$, where $n,m$ are integer numbers. Therefore, all above constants, except two, $\left[j_1\right]_{(\bar{x}\bar{z})^{1+\ep }}$ and $\left[j_4\right]_{(\bar{x}\bar{z})^{-1+\ep }}$, are zero.
The two nonzero coefficients are derived from the leading asymptotics at $\bar{x},\bar{z}\to 0$ of the integrals $j_1$ and $j_4$, respectively. We have
\begin{widetext}
\begin{align}
	\left[j_1\right]_{(\bar{x}\bar{z})^{1+\ep }}&
	=\frac{\Gamma \left(1+\frac\ep2\right)^2}{\Gamma (2+\ep )^2}
	=\frac{e^{-\ep \gamma_E}}{(1+\ep)^2}\left[1-\frac{3 \ep ^2 \zeta _2}{4}+\frac{7 \ep ^3 \zeta _3}{12}+\frac{3}{32} \ep ^4 \zeta _2^2+\ldots\right]
	\, ,\\
	\left[j_4\right]_{(\bar{x}\bar{z})^{-1+\ep }}&
	=2 \Gamma \left(1+\tfrac{\ep }{2}\right) \Gamma (-\ep ) \Gamma \left(-\frac{\ep }{2}\right)
	-\frac{4 \Gamma \left(\frac{\ep }{2}\right)^2 }{\Gamma (1+\ep)^2}\, _3F_2\left(1,\frac{\ep }{2},\frac{\ep }{2};1-\frac{\ep }{2},1+\frac{\ep }{2};1\right)
	\nonumber\\
	&\qquad\qquad=\frac{e^{-\ep \gamma_E}}{\ep^2}\left[-12+11 \ep ^2 \zeta _2-12 \ep ^3 \zeta _3+\frac{19}{40} \ep ^4 \zeta _2^2+\ldots\right]
    \, .
\end{align}
\end{widetext}

\subsection{Solution in terms of Goncharov's polylogarithms}
The perturbative solution of the canonical differential system is expressed in terms of Chen's iterated path integrals
\begin{equation}
	\II(\omega_{i_n},\ldots, \omega_{i_1})=\idotsint\limits_{\boldsymbol{p}\underaccent{C}{>}\boldsymbol{p}_n\underaccent{C}{>}\ldots \underaccent{C}{>}\boldsymbol{p}_1\underaccent{C}{>}\boldsymbol{1}}
	\omega_{i_n}(\boldsymbol{p}_n)\ldots \omega_{i_1}(\boldsymbol{p}_1)\, ,
\end{equation}
where $\boldsymbol{1}=(1,1),\ \boldsymbol{p}=(x',z'),\ \boldsymbol{p_k}=(x_k',z_k')$, $C$ is some contour in $(x',z')$ plane connecting the points $(1,1)$ and $(x',z')$.

Note that $\omega_8$ is singular on the line $z'=x'$ inside the physical region. Therefore, the general solution has a branching locus on this line. However, using physical arguments, we understand that the particular solution determined by the boundary constants has to be analytic at $z'=x'$ on the first sheet. Indeed, we have a posteriori checked this by examining the Frobenius expansion of this specific solution at $z'=x'$.
Therefore the contour $C$ can be arbitrarily deformed provided it does not leave the physical region and has fixed end points $(1,1)$ and $(x',z')$. Such deformations will not alter the special solution.

The possibility to express the iterated path integrals via Goncharov's polylogarithms depends on the possibility to choose a contour and parametrization on it such that the weights become rational functions of the latter. Due to the appearance of algebraic expressions $x_1,\  x_2,\ z_1,\ z_2$, there is no universal choice of the appropriate contour. Note that, in general, each individual integral $\II$ is not path-independent, therefore, we should be very careful. We do it in the following way.

First, by constant transformations of the canonical basis we minimize the appearance of the weights $\omega_{11},\ \omega_{13},\ \omega_{15}$, containing the expression $z_2$, cf.\ eq.~\eqref{eq:x1x2z1z2}. We ensure that these weights appear only in three canonical MIs $J_{12}$, $J_{13}$, and $J_{19}$.
For all other but these three functions, we choose the path to go first along the straight line from $(1,1)$ to $(1,z')$ and then along the straight line from $(1,z')$ to $(x',z')$. For the three remaining functions we choose the path to go first along the straight line from $(1,1)$ to $(x',1)$ and then along the straight line from $(x',1)$ to $(x',z')$.

We express our results in terms of the following functions:
\begin{itemize}
	\item $G(a_1,\ldots |1-x_1)$ with $a_k\in \{0,1,2,1\pm i,1\pm iz_1,1\pm \frac{i}{z_1}\}$.
	\item $G(b_1,\ldots |1-x_2)$ with $b_k\in \left\{0,1,\frac1{z'},\frac{1+z'}{z'},\frac{1+z'}{2z'}\right\}$.
	\item $G(c_1,\ldots |1-x')$ with $c_k\in \left\{0,1,2,1\pm z',1+\frac1{z'}\right\}$.
	\item $G(d_1,\ldots |1-z_2)$ with $d_k\in \left\{0,1,-\frac1{x'},\frac{1-x'}{x'},\frac{1-x'}{2x'}\right\}$.
	\item $G(f_1,\ldots |1-z')$ with $f_k\in \left\{0,1,2,1-x'\right\}$.
\end{itemize}
By using the functional relations between the polylogarithms, we succeed in expressing the results in terms of more convenient functions:
\begin{itemize}
	\item $G(a_1,\ldots |x_1)$ with $a_k\in \left\{0,\pm 1,\pm i,\pm i z_1,\pm\frac{i}{z_1}\right\}$.
	\item $G(b_1,\ldots |x_2)$ with $b_k\in\left\{0,1,-\frac{1}{z'},\frac{z'-1}{2 z'},\frac{z'-1}{z'}\right\}$.
	\item $G(c_1,\ldots |x')$ with $c_k\in \left\{0,\pm 1,\pm z',-\frac{1}{z'}\right\}$.
	\item $G(d_1,\ldots |z_2)$ with $d_k\in\left\{0,1,\frac{1}{x'},\frac{x'+1}{2 x'},\frac{x'+1}{x'}\right\}$.
	\item $G(f_1,\ldots |z')$ with $f_k\in\left\{0,\pm 1,x'\right\}$.
\end{itemize}
The results for the MIs are presented in the supplementary materials in terms of these functions up to transcendental weight 4.

\subsection{Additional relations between master integrals}
Once the solutions for the canonical functions $J_k$ are found, we can search for additional linear relations between MIs which have been missed during the stage of IBP reduction. Indeed, for canonical functions these relations necessarily need to have constant coefficients (independent both on kinematic variables and on $\ep$). Such relations can be easily discovered from a sufficiently deep series expansion in $\ep$.
Subsequently, using differential equations and exact boundary constants, these relations can be strictly proven.
In this way, we can prove two additional identities
\begin{align}
\label{eq:ex-sym}
	&j_9
	=\frac{(1 + \ep)^2x'}{z'}j_1
    -\frac{\bar{x}(1 + z')\ep^2}{4z'}j_2
	\nonumber\\
	&+\frac{(1 + x')\bar{z}(1 + \ep)\ep}{2z'}j_5
	-\frac{(1 + x')(1 + z')\ep^2}{4z'}j_8\, ,\nonumber
	\\
	&j_{16}=j_7\, .
\end{align}
Therefore, we can exclude $j_9$ and $j_{16}$ from the list of MIs.
We are unaware whether these relations follow from IBP identities alone.
For the reader's convenience let us present also the reduction of the integral $j_{8n}=\RR_5(-1,0,1)$ mentioned earlier:
\begin{equation}
    j_{8n}= \frac{z'}{x'}j_8 -\frac{\bar{z}}{x'} j_5\,.
\end{equation}

\section{Method of radial-angular decomposition}
\label{sec:rad-ang}
In this approach, the MIs are evaluated by separating them into radial and angular components, allowing the angular part to be computed independently. The motivation behind this decomposition is to take advantage of the process-independent nature of the angular integrals. These angular integrals can be computed in a general form. Once computed, they can be applied to any specific process by choosing an appropriate reference frame, combining the angular and radial parts to compute phase-space integrals in that chosen frame. The latter is process and kinematics dependent.

The first step is to express the propagators and phase-space integration measures within the chosen reference frame. We then transform the integrals into the van Neerven form~\cite{vanNeerven:1985xr,Ravindran:2002dc,Ravindran:2002na,Ravindran:2003um}, which are subsequently mapped to the coordinate-independent Somogyi representation~\cite{Somogyi:2011ir}.
The Somogyi integrals are then evaluated using their MB representations. After evaluating the angular integrals, we return to the reference frame of interest and complete the final one-fold parametric integral over the radial part.

For the double real emission diagrams at NNLO, we move to the center-of-mass (COM) frame of the two emitted partons having momenta $k_1$ and $k_2$. We have
$$
\mathbf{k}_1+\mathbf{k}_2=0
\, ,
$$
and we parametrize the momenta as\\
\begin{gather}
 p_1=\frac{\left(s-t-q^2\right)}{2 \sqrt{s_{12}}}(1,\mathbf{0}_{d-3},0,0,1)\, ,\nonumber\\
 p_2=\frac{\left(s-s_{12}\right)}{2 \sqrt{s_{12}}}(1,\mathbf{0}_{d-3},0, \sin \psi, \cos \psi)\, ,\nonumber\\
 k_1=\frac{\sqrt{s_{12}}}{2}(1,\mathbf{0}_{d-3}, \sin \phi \sin \theta, \cos \phi \sin \theta, \cos \theta)\, ,\nonumber\\
 k_2=\frac{\sqrt{s_{12}}}{2}(1,\mathbf{0}_{d-3},-\sin \phi \sin \theta,-\cos \phi \sin \theta,-\cos \theta)\, ,
\end{gather}
with
\begin{gather}
\cos \psi=1-\frac{2 s_{12} t}{\left(s-t-q^2\right)\left(s-s_{12}\right)}\, ,\nonumber\\ s_{12}=\left(k_1+k_2\right)^2\, ,\quad s=\left(p_1+q\right)^2\, ,\quad t=2p_1\cdot p_2\, .
\end{gather}
In the $(k_1,k_2)$ reference frame, the integration over constrained phase space can be written as
\begin{widetext}
\begin{multline}
\int[d PS_3]_{z^\prime}
= \frac{1}{(4 \pi)^d} \frac{\left(s-q^2\right)^{3-d}}{\Gamma(d-3) }\intop_0^\pi d \theta(\sin \theta)^{d-3} \intop_0^\pi d \phi(\sin \phi)^{d-4} \nonumber\\
 \times \intop_0^{s-q^2} d t \intop_{q^2 t /\left(s-q^2\right)}^{s-t} d u\left[t\left(s u-q^2(t+u)\right)(s-t-u)\right]^{\frac{d-4}{2}}\delta\left(z^{\prime}-\frac{t}{s-q^2}\right)
\, .
\end{multline}
\end{widetext}
The variable $u$ is defined through $u=2 p_2\cdot q$ and it satisfies $s_{12}=s-t-u$.
Performing the integral over $t$ using $\delta$-function and passing from $u$ to $z$ via $u=s\left(\frac{-x^{\prime} z^{\prime}}{1-x^{\prime}}+z-z^{\prime} z\right)$, we obtain
\begin{align}
\int[d P S_3]_{z^{\prime}}
= \int[d P S_3]_{z^{\prime}}^{r}\int {d\Omega}
\, ,
\end{align}
where the `radial' and `angular' parts are defined as follows
\begin{align}
\label{eq:dps3omega}
    \int[d P S_3]_{z^{\prime}}^r
    =& \frac{ s^{d-3}(1-{z'})^{d-3}{z'}^{\frac{d-4}{2}}}{(4 \pi )^{d}\Gamma (d-3)}\intop_0^1dz
   ((1-z) z)^{\frac{d-4}{2}}
   \, ,\nonumber\\
    \int d\Omega 
    =&   \intop_0^\pi d\theta \sin ^{d-3}\theta \intop_0^\pi  d\phi\sin ^{d-4}\phi
    \,.
\end{align}
The integral over only the phase-space element can be evaluated to
\begin{equation}
   \int[d P S_3]_{z^{\prime}}^{r}{d\Omega} = \frac{2^{7-4 d} \pi ^{2-d} s^{d-3}
   (1-{z'})^{d-3}
   {z'}^{\frac{d}{2}-2}}{\Gamma
   \left(\frac{d-1}{2}\right)^2}\, .
\end{equation}
All the phase-space integrals can be decomposed into angular and radial parts, which can be combined with the corresponding measure to calculate the final integral.
Given the above parametrization, we can explicitly calculate and express all the scalar products appearing in the denominators of the MIs in this parametric basis and then effectively separate the angular part from the radial one. The denominators in eq.~\eqref{eq:D_i} become
\begin{align}
\label{eq:parametrization-prop}
D_1=&\frac{-s(1-z^{\prime})(1-\cos \theta)}{2\left(1-x^{\prime}\right)}, \nonumber\\
D_2 =&-\frac{s\left(x^{\prime}+z+x^{\prime} z\left(-1+z^{\prime}\right)+z^{\prime}-z z^{\prime}\right)}
{2\left(1-x^{\prime}\right)}\nonumber\\
&+\frac{s\left(z^{\prime}-z\left(1+z^{\prime}\right)+x^{\prime}\left(-1+z+z z^{\prime}\right)\right)}{2\left(1-x^{\prime}\right)} \cos \theta\nonumber\\
&+s \sqrt{(1-z) z z^{\prime}} \cos \phi \sin \theta,\nonumber\\
D_3=&-\frac{s(1-z^{\prime})(1+\cos\theta)}{2(1-x')} ,\nonumber\\
D_4 =& -\frac{s\left(x^{\prime}+z+x^{\prime} z\left(-1+z^{\prime}\right)+z^{\prime}-z z^{\prime}\right)}{2\left(1-x^{\prime}\right)}\nonumber\\
&-\frac{s\left(z^{\prime}-z\left(1+z^{\prime}\right)+x^{\prime}\left(-1+z+z z^{\prime}\right)\right)}{2\left(1-x^{\prime}\right)} \cos \theta\nonumber\\
&-s \sqrt{(1-z) z z^{\prime}} \cos \phi \sin \theta,\nonumber\\
D_5=&\frac{s}{2}(1-z+zz^{\prime})+\frac{s}{2}(1-z-zz^{\prime}) \cos \theta\nonumber\\
&+s\sqrt{(1-z) zz^{\prime}} \cos \phi \sin \theta, \nonumber\\
D_6=&\frac{s}{2}(1-z+zz^{\prime})-\frac{s}{2}(1-z-zz^{\prime}) \cos \theta\nonumber\\
&-s\sqrt{(1-z) zz^{\prime}} \cos \phi \sin \theta,\nonumber\\
D_7=&s(1-z^{\prime}) z, \nonumber\\
D_8=&\frac{-s(1-z^{\prime})\left(1-\left(1-x^{\prime}\right) z\right)}{\left(1-x^{\prime}\right)}.
\end{align}

The angular integrals, that appear after decomposing the MIs into radial and angular parts, can be recast into the following general form
\begin{widetext}
	\begin{align}
\Omega_{j_1j_2}(\alpha, \beta, A, B, \Gamma)=\int \mathrm{d} \Omega \frac{1}{\left(\alpha+\beta\cos \theta\right)^{j_1}\left(A+B \cos \theta+\Gamma \sin \theta \cos \phi\right)^{j_2}}\, ,
\end{align}
\end{widetext}
which is called the Neerven parametrization~\cite{vanNeerven:1985xr}.
We extensively use partial fraction decomposition to rewrite all angular integrals into this form systematically.
The differential $d\Omega$, which indicates the integral over angles in spherical polar coordinates, is equal to $[d P S_3]_{z^{\prime}}^\Omega$, as defined in eq.~\eqref{eq:dps3omega}.
To make this transparent, consider the MI
	\begin{align}
\RR_4(1,1,1)=\int  \frac{[d P S_3]_{z^{\prime}}}{\left(k_1-p_1\right)^2\left(k_2-p_1-q\right)^2\left(k_1+k_2-p_1\right)^2}\, .
\end{align}
Using the previous list of scalar products parametrized in $(k_1,k_2)$ frame, as given in eq.~\eqref{eq:parametrization-prop}, we can write this integral as
\begin{gather}
    \int  \frac{[d P S_3]_{z^{\prime}}^r}{r}\int \frac{ d\Omega}{\left(\alpha+\beta\cos \theta \right)\left(A+B \cos \theta +\Gamma \sin \theta  \cos \phi \right)}\, ,
    \label{eq:example}
\end{gather}
with
\begin{gather}
r=\frac{s^3({x'}(-1+z)-z)(-1+{z'})^2}{4(-1+{x'})^2},\
\alpha=1,\
\beta=-1,
\nonumber\\
A=zz'-z-{z'},\ B=z-{z'}+z {z'},\
\Gamma=2\sqrt{(1-z) z {z'}} \, .\nonumber
\end{gather}

There is, however, another coordinate-independent way to represent these angular integrals, called the Somogyi parametrization~\cite{Somogyi:2011ir}, which is given by
\begin{align}
\label{eq:somogyi-rep}
\tilde \Omega_{j_1j_2}({v_{11},v_{22},v_{12}})=\int \mathrm{d} \Omega(n) \frac{1}{\left(v_1 \cdot n\right)^{j_1}\left(v_2 \cdot n\right)^{j_2}}
\end{align}
with $v_{ij}=v_i \cdot v_j$. The integration is over the angular variables of the $d$ dimensional vector $n^\mu$.
In order to find the relations between these two representations, we define the vectors $v_1$ and $v_2$ in $d$ dimensions as
\begin{align}
v_1=\left(1, \mathbf{0}_{d-3}, 0,-\frac{\beta}{\alpha}\right), \quad v_2=\left(1, \mathbf{0}_{d-3},-\frac{\Gamma}{A},-\frac{B}{A}\right).
\end{align}
In this basis for $v_1$ and $v_2$, the vector $n$ can be written as
\begin{align}
n=\left(1, \mathbf{n}_{d-3}, \sin \theta \cos \phi, \cos \theta\right),
\end{align}
where $\mathbf{n}_{d-3}$ denotes entries which can be integrated trivially. With this parametrization, we can relate these two representations through
\begin{align}
{\alpha^m A^n} \Omega_{j_1j_2}(\alpha, \beta, A, B, \Gamma)=\tilde \Omega_{j_1j_2}(v_{11},v_{22},v_{12})\, ,
\end{align}
so that they can be solved via their MB representation.

\subsection{Angular integrals through Mellin-Barnes representation}
\label{ss:cat-ang}
The angular integral, eq.~\eqref{eq:somogyi-rep}, admits the following MB representation:
\begin{align}
\label{eq:MB}
&\tilde \Omega_{j_1j_2}=  \frac{2^{d-j-2} \pi^{\frac{d}{2}-1}}{ \Gamma\left(j_1\right) \Gamma\left(j_2\right) \Gamma(d-j-2)} \nonumber \\
& \times \intop_{-i \infty}^{+i \infty}\left[ \frac{\mathrm{d} z_{1 2}}{2 \pi i} \Gamma\left(-z_{1 2}\right)\left(\frac{v_{1 2}}{2}\right)^{z_{1 2}}\right]
\nonumber \\
&\times\intop_{-i \infty}^{+i \infty}\left[\prod_{k=1}^2  \frac{\mathrm{d} z_{k k}}{2 \pi {i}} \Gamma\left(-z_{k k}\right)\left(\frac{v_{k k}}{4}\right)^{z_{k k}}\right]
\nonumber \\
&\times \left[\prod_{k=1}^2 \Gamma\left(j_k+z_k\right)\right] \Gamma({d}/{2}-1-j-z)\, .
\end{align}
The MB integrals are performed over the variables $z_{kl}$. We further define
\begin{gather*}
z=\sum_{k=1}^2 \sum_{l=k}^2 z_{k l}\, ,\quad z_k=\sum_{l=1}^k z_{l k}+\sum_{l=k}^2 z_{k l}\, ,\quad j=\sum_{k=1}^2 j_k\, .
\end{gather*}
In this context, $z$ represents the aggregate sum of all MB variables. For a given index $k$, the variable $z_k$ accounts for the sum of all terms involving $k$ as one of their indices, with $z_{kk}$ being counted twice.

In general, it is not always possible to solve a multifold MB integral into a closed functional form. However, for our purpose, we only need to solve it as a Laurent series expansion in the dimensional regularization parameter $\ep$. To do such an expansion, we need to make sure that the chosen integration contours of the MB integrals do not cross any poles of the integrand when $\ep \rightarrow 0$. This scenario arises when the real parts of the arguments of the Gamma functions are not all positive. In that case, we need to perform an analytic continuation. There are two ways to achieve this. We can either deform the contour~\cite{Smirnov:1999gc} and account for the residues each time the contour crosses a pole. Alternatively~\cite{Tausk:1999vh,Czakon:2005rk}, we can fix the contour, which is typically chosen parallel to the imaginary axis by separating the poles of $\Gamma(\dots+z_{kl})$ and $\Gamma(\dots-z_{kl})$, and track the movement of the poles of the Gamma function. Each time a pole crosses a contour, the corresponding residue is incorporated into the expression. After that we can safely expand the integrand around $\ep \rightarrow 0$.
This process effectively decomposes the original MB integral into several MB integrals that are finite in $\varepsilon$. We compute these analytically by transforming them into integrals over real variables, with the final results of the angular integrals expressed in terms of multiple polylogarithms. This is discussed in detail in a recent publication~\cite{Ahmed:2024pxr} by some of us.

For the case of two-denominator angular integrals, as given in eq.~\eqref{eq:somogyi-rep}, they can also be represented using hypergeometric functions, Appell functions of the first kind and Lauricella functions~\cite{vanNeerven:1985xr,Somogyi:2011ir,Lyubovitskij:2021ges,Lyubovitskij:2021ges}. The Lauricella functions appear when both momenta, $v_1$ and $v_2$, are massive. Expanding the hypergeometric and Appell functions in Laurent series in $\varepsilon$ shows complete consistency with our results. Double massive integrals, where both $v_1^2\neq 0$ and $v_2^2\neq 0$, can be decomposed into two single-massive, two-denominator integrals employing partial fraction decomposition~\cite{Lyubovitskij:2021ges} which reads
\begin{widetext}
\begin{align}
\tilde\Omega_{j_1, j_2}\left(v_{11}, v_{22}, v_{12}\right) & =\sum_{n=0}^{j_1-1}\binom{j_2-1+n}{j_2-1} \lambda_{ \pm}^{j_2}\left(1-\lambda_{ \pm}\right)^n \tilde\Omega_{j_1-n, j_2+n}\left(v_{11}, v_{13}^{ \pm}\right)\nonumber \\
& +\sum_{n=0}^{j_2-1}\binom{j_1-1+n}{j_1-1} \lambda_{ \pm}^n\left(1-\lambda_{ \pm}\right)^{j_1} \tilde\Omega_{j_2-n, j_1+n}\left(v_{22}, v_{23}^{ \pm}\right) \,,
\end{align}
where
	\begin{align}
	& \lambda_{ \pm}=\frac{v_{12}-v_{11} \pm \sqrt{v_{12}^2-v_{11} v_{22}}}{2 v_{12}-v_{11}-v_{22}}\, , \nonumber\\
	& v_{13}^{ \pm}=\left(1-\lambda_{ \pm}\right) v_{11}+\lambda_{ \pm} v_{12}=\frac{v_{11}\left(v_{22} \pm \sqrt{v_{12}^2-v_{11} v_{22}}\right)-v_{12}\left(v_{12} \pm \sqrt{v_{12}^2-v_{11} v_{22}}\right)}{v_{11}+v_{22}-2 v_{12}}\, ,\nonumber \\
	& v_{23}^{ \pm}=\left(1-\lambda_{ \pm}\right) v_{12}+\lambda_{ \pm} v_{22}=\frac{v_{22}\left(v_{11} \mp \sqrt{v_{12}^2-v_{11} v_{22}}\right)-v_{12}\left(v_{12} \mp \sqrt{v_{12}^2-v_{11} v_{22}}\right)}{v_{11}+v_{22}-2 v_{12}}\, .
	\end{align}
\end{widetext}

The advantage of using the MB representation for angular integrals lies in its scalability, making it well-suited for handling integrals with an increasing number of denominators~\cite{Ahmed:2024pxr}. This approach allows for a more systematic treatment of complex integrals, offering a powerful method for solving problems that would otherwise be difficult to tackle analytically.
The results for all angular integrals relevant to the NNLO SIDIS
can be expressed in terms of the following functions:
\begin{widetext}
\begin{itemize}
	\item For two-denominator massless case: $G(a_1,\ldots |1)$ with $a_k\in \left\{0,-\frac{2}{v_{12}-2},\frac{2}{v_{12}},\frac{v_{12}}{v_{12}-2}\right\}$\, .
 \item For one-denominator massive case:\\ $G(b_1,\ldots |1)$ with $b_k\in \left\{0,\frac{1}{2}-\frac{1}{2 \sqrt{1-v}},\frac{1}{2}
   \left(\frac{1}{\sqrt{1-v}}+1\right),\frac{1-\sqrt{1-v}}{\sqrt{1-v}+1}\right\}$, where $v$ can be either $v_{11}$ or $v_{22}$\, .
   \item For two-denominator single massive case: $G(c_1,\ldots |1)$ with $c_k\in \left\{0,\frac{1}{1-\frac{v_{11}}{v_{12}^2}},\frac{1}{\frac{v_{12}}{\sqrt{1-v_{11}}-1}+1},\frac{1}{1-\frac{v_{12}}{\sqrt{1
   -v_{11}}+1}}\right\}$.
\end{itemize}
\end{widetext}
Since two denominator double massive case can be written in terms of single massive cases, the encountered functions are not repeated here. However, in the ancillary file we combine the angular part with the parametric part and provide the results only in terms of one-fold integrals of classical polylogarithms.

\subsection{Resolution of singularities in integrals over z}
\label{sec:parametric}
The singularities arising from the angular integration are of  collinear in nature, while the soft divergence originates from the integration over  $z$. Since the integrand may contain singularities in the integration domain, it is not sufficient to expand the integrand in $\varepsilon$ naively. Instead, we must first isolate the poles before performing the expansion. In what follows, we outline the method used to systematically extract these poles, ensuring that the expansion is carried out correctly.

Consider a function $f(z)$, which is singular at $z \rightarrow a$. We write this function as
\begin{align}
& f(z)=g(z) R(z)\, ,
\end{align}
where $R(z)$ is regular at $z\rightarrow a$ and $g(z)$ captures the singularities. Now we define $\widetilde R$ through
\begin{align}
\widetilde{R} \equiv \lim _{z \rightarrow a}R(z)\, .
\end{align}
Then the integral over $z$ can be rewritten as
\begin{align*}
\int_0^1 f(z) dz & =\int_0^1[g(z) R(z)-\tilde{R} g(z)] dz +\int_0^1 \tilde{R} g(z) dz\, ,
\end{align*}
where we can safely expand the first integrand on the right-hand side. The second term is usually much simpler and can be integrated directly. For example, the integral $\RR_8\left(1,1,1\right)$ 
contains the parametric term,
\begin{align}
   P'_z= \frac{4^{- \varepsilon-3} \pi ^{- \varepsilon-4} (x'-1)^2 s^{ \varepsilon-2} ((1-z)
   z)^{\frac{\varepsilon}{2}} (1-z')^{ \varepsilon-1} z^{\prime\frac{\varepsilon}{2}}}{z \Gamma (1+
   \varepsilon) (x' z (z'-1)+x'-z z'+z+z')}\, .
\end{align}
Since the $z$ independent factors are not of our concern we can just focus on the $z$-dependent part which we denote by
\begin{align}
   P_z=\frac{((1-z) z)^{\frac{\varepsilon}{2} }}{z \left(z \left(-x'\right)+z x' z'+x'-z z'+z'+z\right)}.
\end{align}
The Laurent series expansion of $P_z$ in powers of $z$ contains $z^{-1+\frac{\ep}{2}}$ in the leading term. It indicates that $P_z$ has a singularity at $z\rightarrow 0$, which is regulated by $\ep$.
The corresponding angular part does not contain any singularity at $z\rightarrow0$. Therefore, we can safely use the $\ep$ expanded expression of the angular integral.
We expand the angular part to $\ep^3$ and denote it by $\Omega_z$. To evaluate the $z$-integration of $f(z)=P_z \Omega_z$, we have to identify $g(z)$ and $R(z)$, as mentioned above. It is clear from the expression that we can choose $g(z)=z^{-1+\frac{\ep}{2}}$. Hence, the corresponding regular function is $R(z)=P_z \Omega_z z^{1-\frac{\ep}{2}}$, which is finite at $z\rightarrow0$. Being regular, we can now expand $R(z)$ around $z\rightarrow0$ and determine $\widetilde{R}$. The term $(g(z) R(z)-\tilde{R} g(z))$ can be expanded in $\varepsilon$ safely and the integral $\int_0^1 \tilde{R} g(z) d z$ is a simple integration that can be directly evaluated.

A more complex scenario arises when the angular part itself exhibits singularities within the $z$-integration domain. In particular, this occurs primarily in the massless case. The challenge here is that if we have an $\ep$-expanded expression for the angular part up to a finite order, we cannot directly use it due to the same limitation encountered when expanding the integrand containing singularities before extracting the poles.
To address this issue, we must isolate the singular component from the all-order expansion of the angular integral. Given that the $z$-integration captures the soft singularity, we hypothesize, as described in refs.~\cite{Ahmed:2024pxr,Wunder:2024btq}, that the all-order singularity from the $z$-integration of the angular part takes the form $z^{n_1 + \frac{n_2}{2}\varepsilon}$, where $n_1$ and $n_2$ are integers. After correctly factoring out this soft singularity, it can be combined with the singularity from the parametric part, allowing us to apply the previous method for obtaining the correct form of the integrand.
For example, consider the integral $\RR_2(1,1,1)$, where this approach is necessary to correctly manage the singularities. 
The angular part of the integral in terms of logarithms and polylogarithms reads
\begin{widetext}
	\begin{align}
\Omega_z^{\prime}&=  \frac{2\pi}{\varepsilon z}\left(z^\prime +z - z z^\prime\right)+\frac{\pi}{z}\left(z^\prime +z - z z^\prime\right) \ln \left(\frac{z}{z^\prime+z-zz^\prime}\right) \nonumber\\
& +\frac{\pi \varepsilon}{4z}\left(z^\prime +z - z z^\prime\right)\left\{2 \operatorname{Li}_2\left(\frac{z}{z z^\prime - z^\prime-z}+1\right)+\ln ^2\left(\frac{z}{-zz^\prime+z^\prime+z}\right)\right\}\nonumber \\
& +\frac{\varepsilon ^ { 2 }}{24 z} \pi \left(z^\prime +z - z z^\prime\right)  \left\{-6 \operatorname{Li}_3\left(\frac{z}{-z^\prime z+z+z^\prime}\right)-6 \operatorname{Li}_3\left(\frac{z}{z z^\prime-z^\prime-z}+1\right)\right.\nonumber \\
& +\ln ^3\left(\frac{z}{-zz^\prime+z^\prime+z}\right)-3 \ln \left(\frac{z}{z z^\prime - z^\prime-z}+1\right) \ln ^2\left(\frac{z}{-zz^\prime+z^\prime+z}\right)\nonumber \\
& +\left.\pi^2 \ln \left(\frac{z}{-zz^\prime+z^\prime+z}\right)+6 \zeta(3)\right\}+\frac{\pi \varepsilon^3}{8z}\left(z z^\prime - z^\prime -z\right) \operatorname{Li}_4\left(\frac{z z^\prime - z^\prime}{z}\right)+O\left(\varepsilon^4\right)\, .
\end{align}
\end{widetext}
By analyzing the expression, we factor out $z^{-1+\frac{\varepsilon}{2}}$ to extract the all-order soft singularity, allowing us to redefine the angular component as $\Omega_z = \Omega'_z z^{1-\frac{\varepsilon}{2}}$. The parametric portion of the integral is given by
\begin{align}
    P^\prime_z=\frac{(1-{x^\prime})  \pi ^{-\varepsilon -4} s^{\varepsilon
   -2} ((1-z) z)^{\varepsilon /2} (1-{z^\prime})^{\varepsilon +1}
   {z^\prime}^{\varepsilon /2}}{4^{3+\varepsilon}(z-1) ({z^\prime}-1)^2 ({z^\prime}-z
   ({z^\prime}-1)) \Gamma (\varepsilon +1)}\, .
\end{align}
To counterbalance the additional factor of $z^{1-\frac{\varepsilon}{2}}$ introduced in the angular part, we multiply the parametric part by $z^{-1+\frac{\varepsilon}{2}}$, effectively adjusting the overall expression. We then define the $z$-dependent factor of the parametric part as
\begin{equation}
    P_z=\frac{z^{\frac{\varepsilon }{2}-1} (-((z-1) z))^{\varepsilon /2}}{(z-1) (z
   {z^\prime}-z-{z^\prime})}.
\end{equation}
Here we observe that there are two different poles arising when $z\rightarrow0$ or $z\rightarrow1$. It means we have to provide two separate counter-terms to resolve the singularities. As before, we express $f(z) = P_z \Omega_z$ and identify $g_1(z) = z^{-1 +\varepsilon}$, leading to $R_1(z) = P_z \Omega_z z^{1 -\varepsilon}$. Expanding $R_1(z)$ around $z \to 0$ and extracting the zeroth-order term, we obtain
\begin{widetext}
	\begin{multline}
    \widetilde{R}_1 =-\frac{1}{180} \varepsilon ^3 \left(-\frac{15}{16} \pi  \ln
   ^4\left({z^\prime}\right)-\frac{15}{8} \pi ^3 \ln
   ^2\left({z^\prime}\right)-\frac{7 \pi ^5}{16}\right)-\frac{1}{180} \varepsilon^2 \left(\frac{15}{2} \pi  \ln ^3\left({z^\prime}\right)+\frac{15}{2} \pi
   ^3 \ln \left({z^\prime}\right)\right)\\
   -\frac{1}{180} \varepsilon  \left(-45 \pi
    \ln ^2\left({z^\prime}\right)-15 \pi ^3\right)-\pi  \ln
   \left({z^\prime}\right)+\frac{2 \pi }{\varepsilon }\, ,
\end{multline}
and
\begin{multline}
   \int_0^1 \widetilde{R}_1 g_1(z) d z = \frac{2 \pi }{\varepsilon ^2}-\frac{\pi  \ln \left({z^\prime}\right)}{\varepsilon
   }+\frac{1}{12} \pi  \left(3 \ln ^2\left({z^\prime}\right)+\pi
   ^2\right)-\frac{1}{24} \varepsilon  \left(\pi  \ln \left({z^\prime}\right)
   \left(\ln ^2\left({z^\prime}\right)+\pi ^2\right)\right)\\
   +\frac{\pi
   \varepsilon ^2 \left(15 \ln ^4\left({z^\prime}\right)+30 \pi ^2 \ln
   ^2\left({z^\prime}\right)+7 \pi ^4\right)}{2880}+O\left(\varepsilon ^3\right)\, .
\end{multline}
\end{widetext}
Next, we identify $g_2(z)=(1-z)^{-1+\frac{\varepsilon}{2}}$, leading to $R_2(z)=P_z\Omega_z (1-z)^{1- \frac{\varepsilon}{2}}$. Expanding $R_2(z)$ around $z \to 1$ and extracting the zeroth-order term, we obtain
\begin{equation}
    \widetilde{R}_2 = \frac{2\pi}{\varepsilon}\, .
\end{equation}
and
\begin{equation}
    \int_0^1 \widetilde{R}_2 g_2(z) d z = \frac{4 \pi }{\varepsilon ^2}+O\left(\varepsilon ^3\right)\, .
\end{equation}
By safely expanding $[g_1(z) R_1(z) +g_2(z)R_2(z) - \widetilde{R}_1 g_1(z) -\widetilde{R}_2 g_2(z)]$ in terms of $\varepsilon$ and incorporating the pole terms, we arrive at a well-suited integrand that can be easily handled via numerical integration.

It should be noted that not all MIs require singularity resolution, and for those, the highest poles are of order $\varepsilon^{-1}$ ($\RR_1(0,0,1)$, $\RR_5(1,1,1)$, $\RR_6(1,1,1)$, $\RR_7(0,1,1)$, $\RR_{10}(1,1,1)$, $\RR_{12}(1,1,1)$, $\RR_{13}(1,1,1)$) or $\varepsilon^0$ ($\RR_3(0,0,1)$, $\RR_5(0,0,1)$, $\RR_5(0,0,2)$, $\RR_5(0,1,1)$, $\RR_6(0,1,1)$). But the integrals where resolution of singularities is required have the highest pole of order $\varepsilon^{-2}$ ($\RR_1(1,1,1)$, $\RR_2(1,1,1)$, $\RR_3(1,1,1)$, $\RR_4(1,1,1)$, $\RR_7(1,1,1)$, $\RR_8(1,1,1)$, $\RR_{11}(1,1,1)$).

A key feature of this analysis is the systematic separation of soft and collinear singularities. Additionally, the angular components are independent of the experimental observables, making this technique applicable to a wide range of processes. The process-dependent contributions appear only in the parametric integrals, which become numerically straightforward once the singularities are resolved.

Although the complete results in this method are provided in terms of one-fold integrals over classical polylogarithms, one may attempt to evaluate the parametric integrals fully analytically for the sake of elegance. In appendix~\ref{app:int-z}, we illustrate how one can proceed through an example.

We further cross-check the relations presented in eq.~\eqref{eq:ex-sym}. Notably, an interesting feature of eq.~\eqref{eq:ex-sym} is that the angular component of $j_7$ is entirely massless, whereas the angular component of $j_{16}$ is massive. Despite this difference, the corresponding phase-space integrals are found to be equal.

\subsection{Results of the method of radial-angular decomposition}
\label{sec:res}
The angular integrals, expressed in terms of Goncharov polylogarithms or classical polylogarithms, are combined with the parametric part following the method outlined in section~\ref{sec:parametric}. These parametric integrals are one-dimensional integrals over $z$ from 0 to 1. While they can be evaluated analytically using the GPL algebra, they can also be computed numerically. Even when the parametric integral is computed numerically, we can fully extract the singularities in $\varepsilon$ analytically, which ensures that we can observe the cancelation of soft and collinear singularities in the NNLO SIDIS cross section analytically.
Both methods, numerical and analytical, produce results that are in high agreement with high precision. We further validate the full phase-space integrals using differential equations, as discussed in section~\ref{sec:de}.

Decomposing the integrals into angular and parametric parts provides deeper insight into the singularity structure, clearly identifying the sources of soft and collinear singularities. An additional benefit of this method is that the angular integrals are evaluated in a general form, removing the need to compute each angular MI individually, aside from variable substitutions. Only the parametric parts depend on the specific process. We calculate the MIs up to $O(\varepsilon^2)$.

\begin{widetext}
\section{Description of the supplementary materials}
We provide the results of the present paper in the files:
\begin{itemize}
    \item \verb|de_results.m| --- results of the differential equations method in the \textit{Mathematica}-readable file. The file format is \verb|{RRtoJ,JtoG,Grules,notations}|, where
    \begin{itemize}
        \item \verb|RRtoJ|  --- rules expressing $\RR_k(n_1,n_2,n_3)$ via canonical basis $J_k$,
        \item \verb|JtoG| --- rules expressing $J_k$ via \verb|G1,G2,...,G1627| functions,
        \item \verb|Grules| --- definition rules for \verb|G1,G2,...,G1627| functions as GPLs
        \item \verb|notations| --- definition of notations \verb|x1,z1,x2,z2|.
    \end{itemize}
    \item \verb|rad_results.m| --- results of the radial-angular decomposition method in the \textit{Mathematica}-readable file. The file format is \verb|{RRtoInt,GtoLi,prefactors,notations}|, where
    \begin{itemize}
        \item \verb|RRtoInt| ---  integrands in symbolic form containing GPLs and classical polylogarithms,
        \item \verb|GtoLi| --- rules to convert the GPLs to classical polylogarithms,
        \item \verb|prefactors| --- expression for prefactors in the integrands,
        \item \verb|notations| --- two square roots present in the expression.
    \end{itemize}
    \item \verb|Evaluation.nb| --- \textit{Mathematica} notebook demonstrating the usage of the two above files.
\end{itemize}
\end{widetext}

\section{Conclusion}
In this article, we compute the NNLO phase-space integrals relevant to SIDIS. By applying the Cutkosky rules, we convert phase-space integrals into loop integrals, enabling the use of IBP identities to reduce them to a set of MIs. These MIs are then evaluated using two distinct methods.

The first, the differential equation method, uses the reduction of the differential equation system to $\ep$-form which allows one to construct a canonical basis and to solve the integrals efficiently and systematically. In particular, this approach enabled us to uncover nontrivial relations between the integrals that do not appear to be derivable from IBP reduction.

The second, the radial-angular decomposition method, offers the advantage of separating soft and collinear singularities, providing deeper insight into the singularity structure of the process. In addition, we develop a general approach to resolve singularities necessary to combine the radial part with the angular one. This work serves as a foundational framework for similar processes and as a guide for further precision calculations.

\section*{Acknowledgements}
The work of RL is supported by the Russian Science Foundation, grant no. 20-12-00205. The work of TA and AR is supported by Deutsche Forschungsgemeinschaft (DFG) through the Research Unit FOR2926, \textit{Next Generation perturbative QCD for Hadron Structure: Preparing for the Electron-ion collider},
project number 409651613. Part of the work of SMH is completed during the visiting period in the University at Buffalo. SMH would like to thank Dr. Doreen Wackeroth for useful discussion.

\section*{Note added}
When this work was finished, we have learned about Ref. \cite{Bonino:2024adk} where the same set of integrals was calculated in terms of classical polylogarithms up to transcendental weight 3. We have performed a numerical comparison of our results with those of Ref. \cite{Bonino:2024adk} and found a perfect agreement.

\appendix

\section{Analytic integration over z}
\label{app:int-z}
Let us consider the integral $\RR_4(1,1,1)$ up to $O\left(\varepsilon ^0\right)$, which can be written as
\begin{widetext}
	\begin{align}
\RR_4(1,1,1) &= C_4 \times \int_0^1 d z\left[\frac{2 \pi }{\varepsilon ^2 {z^\prime}}+\frac{\pi}{\varepsilon
   {z^\prime}}\left\{\frac{2-2 {x^\prime}}{z {x^\prime}-z+1}-G\left(\frac{1}{1-{z^\prime}},1\right)\right\}\right. \nonumber\\
& +\frac{\pi}{12 z^{\prime}}\left\{3 G\left(\frac{1}{1-{z^\prime}},1\right)^2+\frac{12}{z}G\left(\frac{1}{1-{z^\prime}},1\right)+\frac{24}{z}G\left(\frac{1}{1-z},1\right)\right. \nonumber\\
& +\frac{-12 G\left(\frac{z \left(-{z^\prime}\right)+z-1}{z-1},1\right)-12 G\left(\frac{1}{(z-1) z+1},1\right)+\pi ^2 z \left(z
   {x^\prime}-z+1\right)}{z \left(z {x^\prime}-z+1\right)}\Bigg\} \nonumber\\
& +O\left(\varepsilon\right)\Big],
\end{align}
\end{widetext}

where
\begin{align}
    C_4=-\frac{4^{-\varepsilon -3} \pi ^{-\varepsilon -4} \left({x^\prime}-1\right){}^2 s^{\varepsilon -2} \left(1-{z^\prime}\right){}^{\varepsilon } {z^\prime}^{\varepsilon
   /2}}{\left({z^\prime}-1\right) \Gamma (\varepsilon +1)}.
\end{align}
To analytically evaluate this integral, the variable $z$ must first be shifted to the right-most argument of the GPLs~\cite{Vollinga:2004sn}. To facilitate this, we employ the feature ``fibration bases" of {\tt PolyLogTools}~\cite{Duhr:2019tlz} to achieve this goal. After performing the integration, we get
\begin{widetext}
	\begin{multline}
\RR_4(1,1,1) = C_4\times\bigg[\frac{2 \pi }{\varepsilon ^2 {z^\prime}}-\frac{\pi }{\varepsilon
   {z^\prime}} \left\{2 G\left(\frac{1}{1-{x^\prime}},1\right)+G\left(\frac{1}{1-{z^\prime}},1\right)\right\} +\frac{\pi}{12 {z^\prime}}\bigg\{12 G\left(\frac{1}{1-{x^\prime}},1\right) G\left(\frac{1}{1-{z^\prime}},1\right)\\
   +12 G\left(\frac{1}{1-{x^\prime}},\frac{1}{1-{z^\prime}},1\right)+24
   G\left(0,\frac{1}{1-{x^\prime}},1\right)-12 G\left(\frac{1}{1-{x^\prime}},1,1\right)+6 G\left(0,0,{z^\prime}\right)-12
   G\left(0,\frac{1}{1-{z^\prime}},1\right)-\pi ^2\bigg\}\bigg].
\end{multline}
\end{widetext}
The higher order terms of this integral are omitted here because of  the large size of the expression, and no new techniques are involved in their evaluation. This result agrees with the expression obtained in section~\ref{sec:de} via differential equations method.

An interesting feature appears when square roots are involved. For example, the integrand for $\RR_5(1,1,1)$ has square roots over a function of $z$ in the weight arguments. One such GPL at $O(\varepsilon)$ is
\begin{align*}
G\left(0,\frac{\xi -z \left(x^{\prime}-1\right) \left(z^{\prime}-1\right)+x^{\prime} z^{\prime}+1}{\xi -x^{\prime} z^{\prime}+z \left(x^{\prime}-1\right)
   \left(z^{\prime}+1\right)+1},1\right),
\end{align*}
where
\begin{widetext}
	\begin{align*}
     \xi=\sqrt{z^2 \left(x^{\prime}-1\right){}^2 \left(z^{\prime}-1\right){}^2-2 z \left(x^{\prime}-1\right) \left(z^{\prime}-1\right) \left(x^{\prime}
   z^{\prime}+1\right)+\left(x^{\prime} z^{\prime}-1\right){}^2}.
\end{align*}
\end{widetext}
In fact, at all orders of $\varepsilon$, the same square root structure appears, with only the weight of the GPLs increasing. This implies that no GPL will involve two distinct types of square roots. In other words, these square roots can be rationalized \cite{Besier:2018jen} through transformation of variable, as the argument inside the square root is quadratic in $z$.
This observation is in agreement with the differential equations approach of section~\ref{sec:de}, which allowed us to obtain the results for all MIs in terms of GPLs.

The next step is to move the transformed variable to the rightmost argument of the GPL. Unlike the previous example, this is not straightforward. The technique involves first reducing the weight by differentiation, then shifting the transformed variable to the right, and reconstructing the original expression through iterative integration. This process can be handled algorithmically, as described in refs.~\cite{Gehrmann:2001jv,DelDuca:2009ac,DelDuca:2010zg,Bolzoni:2009ye}. We compute the analytic expression using our in-house code and verify the results numerically.
Additionally, we observe that the single square root feature persists across all the MIs.

\bibliographystyle{JHEP}
\bibliography{PSmasters}
\end{document}